\documentclass[conference]{IEEEtran}

\usepackage{graphicx}
\usepackage{xcolor}
\usepackage[caption=false,font=footnotesize]{subfig}

\usepackage{array, amsmath, amssymb, amsfonts, bm}
\usepackage{mathrsfs}
\usepackage{multirow, booktabs}
\usepackage{algorithm, algpseudocode}
\usepackage{xargs}
\usepackage{hyperref}

\usepackage{caption}
\usepackage{comment}
\usepackage{kantlipsum}
\usepackage{microtype}
\usepackage[normalem]{ulem}
\usepackage[style=ieee, 
            citestyle=numeric-comp,
            maxnames=2, 
            minnames=1, 
            doi=false,
            isbn=false,
            url=false,
            date=year,
            ]{biblatex}
            
\useunder{\uline}{\ul}{}

\newcommand{\grey}[1]{\textcolor{gray}{#1}}

\newcommand{\diag}{\mathrm{diag}}

\newcommand{\setR}{\mathbb{R}}
\newcommand{\setRp}{\mathbb{R}_+}
\newcommand{\setC}{\mathbb{C}}
\newcommand{\setSp}{\mathbb{S}_+}

\newcommand{\eye}{{\bf I}}

\newcommand{\adj}{\mathsf{H}}

\newcommand{\distnormal}[2]{\mathcal{N}\left({#1}, {#2}\right)}
\newcommand{\distcmpnormal}[2]{\mathcal{N}_{\mathbb{C}}\left({#1}, {#2}\right)}
\newcommand{\distcmpstudent}[2]{\mathcal{T}_{\mathbb{C}}^{\nu}\left({#1}, {#2}\right)}
\newcommand{\distcmplepto}[2]{\mathcal{GG}_{\mathbb{C}}^{\beta}\left({#1}, {#2}\right)}

\NewDocumentCommand\newletter{m m o m m}{%
\NewDocumentCommand#1{s t@ o}{%
\IfBooleanTF{##1}{\mathbf{\MakeUppercase{#2}}\IfValueT{#3}{^{#3}}}{%
\IfBooleanTF{##2}{\mathbf{#2}\IfValueT{#3}{^{#3}}_{\IfValueTF{##3}{##3}{#5}}}{%
{#2}\IfValueT{#3}{^{#3}}_{\IfValueTF{##3}{##3}{#4}}%
}}}}

\NewDocumentCommand\newletterbm{m m o m m}{%
\NewDocumentCommand#1{s t@ o}{%
\IfBooleanTF{##1}{\bm{\MakeUppercase{#2}}\IfValueT{#3}{^{#3}}}{%
\IfBooleanTF{##2}{\bm{#2}\IfValueT{#3}{^{#3}}_{\IfValueTF{##3}{##3}{#5}}}{%
{#2}\IfValueT{#3}{^{#3}}_{\IfValueTF{##3}{##3}{#4}}%
}}}}

\newletter{\x}{x}{ftm}{ft}

\newletterbm{\psd}{\lambda}{nft}{ft}

\newcommand{\scm}[1][nf]{\mathbf{H}_{#1}}

\newcommand{\Q}[1][f]{\mathbf{Q}_{#1}}
\newletter{\q}{q}{fmm}{fm}
\newletter{\xt}{\tilde{x}}{ftm}{ft}
\newletter{\yt}{\tilde{y}}{nftm}{nft}

\newletter{\z}{z}{ntd}{nt}
\newletter{\zs}{z}[*]{ntd}{nt}
\newletter{\msk}{u}{nt}{t}

\newletter{\g}{w}{nfm}{nf}

\newcommand{\sv}[1][nf]{\mathbf{a}_{#1}}

\newcommand{\dec}[1][\theta,f]{g_{#1}}

\newletter{\src}{s}{nft}{nf}

\title{Neural Multichannel Distant Speaker Diarization \\ With Heavy-tailed Source Separation Model
}

\author{
\IEEEauthorblockN{
Sicheng Mao\IEEEauthorrefmark{1} \ \ 
Baihan Li\IEEEauthorrefmark{1}\IEEEauthorrefmark{2} \ \ 
Mathieu Fontaine\IEEEauthorrefmark{1}\ \ 
Anthony Larcher\IEEEauthorrefmark{3} \ \ 
Roland Badeau\IEEEauthorrefmark{1}
}
\IEEEauthorblockA{
\IEEEauthorrefmark{1} LTCI, Telecom Paris, Institut Polytechnique Paris, France\\
\IEEEauthorrefmark{2} SJTU Paris Elite Institute of Technology, Shanghai Jiao Tong University, China \\
\IEEEauthorrefmark{3} LIUM, Le Mans University, France\\
}
}

\begin{document}
\maketitle
\begin{abstract}
Distant speaker diarization remains challenging due to difficult acoustic environments, varying numbers of speakers and overlapping speech. Model-driven methods are proposed to exploit the speech source features in multi-channel recordings that help diarization. This paper generalizes a neural model that jointly learns to perform blind source separation and diarization over speech mixtures (neural FCASA) with heavy-tailed models. The popular Gaussian distribution has been applied for variance modeling in the original source separation model, which we replace with two families of heavy-tailed models (Leptokurtic Generalized Gaussian distribution and Student's t distribution) to better capture the heavy-tailedness in speech signals. Thanks to the Gaussian scale mixture model, we are able to unify the proposed method and the original one under the same form of learning objective. 
Our experiments show consistent large improvements in Diarization Error Rate (DER) and Jaccard Error Rate (JER) compared to the baseline on various corpora.

\end{abstract}
\section{Introduction}
\label{sec:intro}

Speaker diarization (SD) addresses the question of ``who spoke when'' by logging speaker-specific salient events in audio data.
Distant speaker diarization remains challenging due to difficult acoustic environments such as meeting scenarios~\cite{pardoSpeakerDiarizationMultiple2006,watanabeCHiME6ChallengeTackling2020,cornellCHiME8DASRChallenge2024}, where speech signals are often recorded by a distant microphone array and contain more noise and reverberation, with a varying number of speakers and overlapping speech. 

To enhance the robustness of a distant diarization system, data-driven paradigm performs data augmentation by noise, reverberation and synthesized overlapping speech and trains powerful speaker feature extractors~\cite{snyderDeepNeuralNetwork2017} within pipeline systems~\cite{landiniBayesianHMMClustering2022} or end-to-end neural diarization (EEND) systems~\cite{fujitaEndEndNeuralSpeaker2019, horiguchiEncoderDecoderBasedAttractors2022}. This paradigm recently drives the application of Large Language Models (LLMs)~\cite{wangDiarizationLMSpeakerDiarization2024, yinSpeakerLMEndtoEndVersatile2025} that demonstrate the scaling laws on the diarization task.

On the other hand, model-driven methods still hold certain interest by incorporating traditional signal processing knowledge, especially for multi-channel distant speech recordings. The direction-of-arrival (DOA), the inter-channel time delay (ITD) and other distant speech specific features captured by multi-channel recordings can be used to characterize the speech sources and hence contribute to multi-channel diarization tasks. \cite{angueraAcousticBeamformingSpeaker2007, mariotteASoBOAttentiveBeamformer2024} incorporate beamforming models and extract the steering vector of a signal to enhance the speech features.  The multi-channel EEND \cite{horiguchiMultiChannelEndtoEndNeural2022} adapts the Transformer encoder to extract spatio-temporal information and the neural FCASA~\cite{bandoNeuralBlindSource2024} incorporates a source separation model that performs joint speaker separation and diarization.
\begin{figure}[t]
    \centering \includegraphics[width=\linewidth]{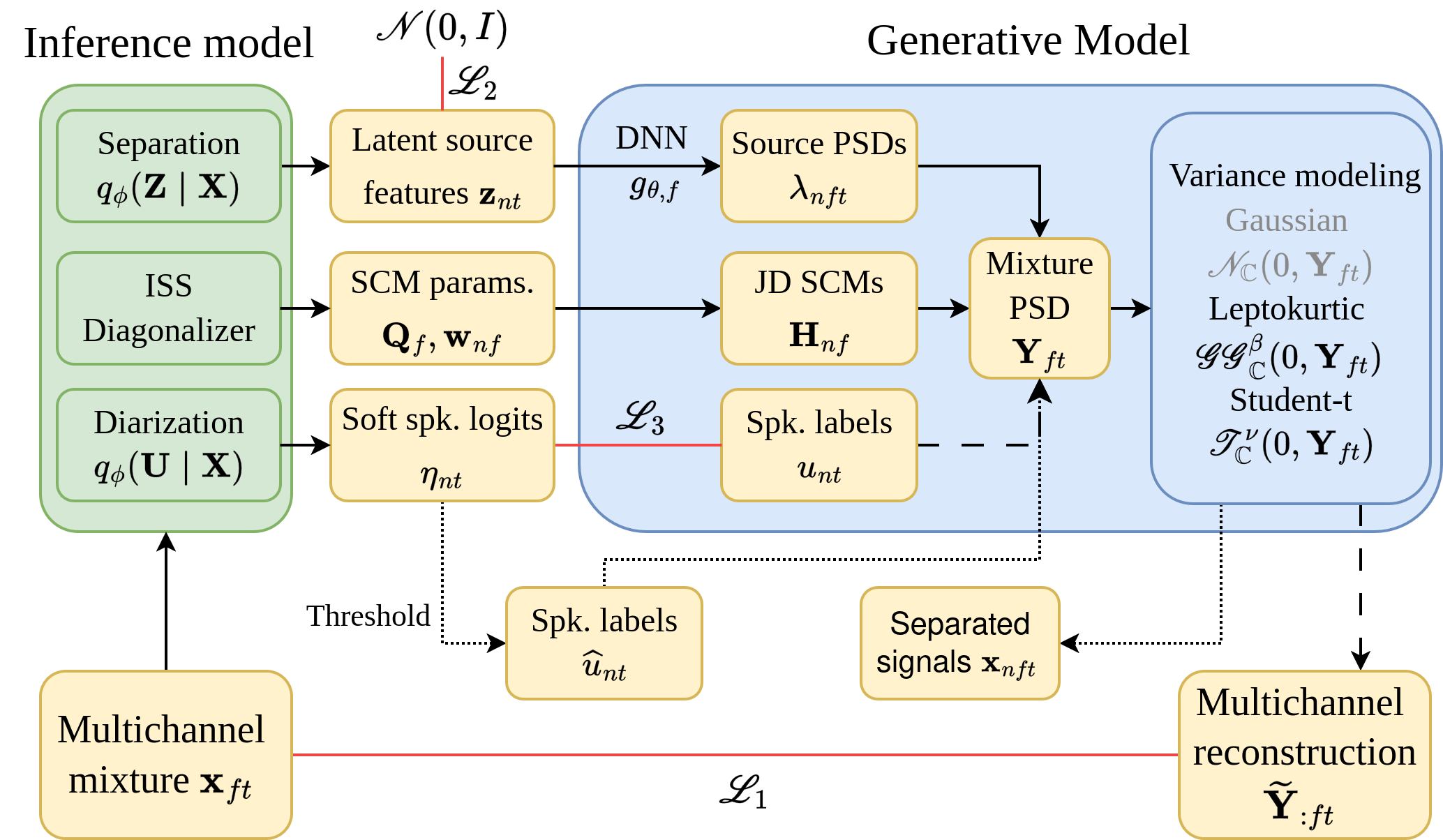}
    \caption{The overview of our model, training and inference}\small{Dashed: training pipeline; Dotted: inference pipeline; Solid: both; Red: loss terms. The replaced Gaussian model is faded in gray.}
    \label{fig:placeholder}
\end{figure}

While the neural FCASA focuses more on enhancing the source separation performance with diarization information, from a multi-task learning perspective, it is also beneficial to the diarization system with an incorporated separation model, because, intuitively, a well-separated speech mixture would reduce the speaker diarization to a simpler voice activity detection task. Therefore, we hypothesize that a better separation model would benefit the diarization performance more. 
It is known that heavy-tailed separation models~\cite{fontaineGeneralizedFastMultichannel2022} are more robust to a wide range of difficult scenarios (e.g. noisy conditions, dynamic range, reverberation etc.). Hence, in this work, we replace the Gaussian model in neural FCASA by heavy-tailed distributions from two typical families: the Leptokurtic Generalized Gaussian distribution and the Student's t distribution. We then train models and conduct a thorough evaluation of diarization performance over various corpora including AMI ~\cite{carlettaAMIMeetingCorpus2005}, AliMeeting~\cite{yu2022alimeeting} and CHiME-6~\cite{watanabeCHiME6ChallengeTackling2020} across different setups and parameter settings.  Our results show that this method outperforms the baseline by a large margin on these corpora.

The main contribution of this work is to propose a neural speaker diarization method with a heavy-tailed separation model as an improvement and generalization of the neural FCASA.

\section{Blind source separation and diarization in the neural FCASA}
This section describes the formulation of 
the neural FCASA, which we further extend with heavy-tailed separation models.


\subsection{Generative model of multichannel mixture signal}
Let us denote $\x@ \in \setC^M$ as an $M$-channel mixture signal, which is a sum of $N$ image signals $\mathbf{x}_{nft}$ generated from a source signal $\src \in \setC$, represented in the Short Time Fourier Transform (STFT) domain.
The neural FCASA has been proposed to jointly train the separation and diarization by introducing a source activity temporal mask $\msk \in \{0,1\}$ to the typical blind source separation model~\cite{duongUnderDeterminedReverberantAudio2010, bandoNeuralFastFullRank2023}: 
\begin{align}
  \x@ = \sum_{n=1}^N \msk\mathbf{x}_{nft} = \sum_{n=1}^N \msk\sv \src, \label{eq:spk-tf-mixture}
\end{align}
where $\sv \in \setC^M$ is the steering vector for source $n$, and $t=1,\ldots, T$ and $f=1,\ldots, F$ represent the time and frequency indices, respectively. 

Following the typical variance modeling~\cite{vincentProbabilisticModelingParadigms2011}, each $\src$ is assumed to independently follow a circularly-symmetric complex Gaussian distribution with its power spectrum density (PSD) $\psd \in \setRp$ as the variance of the Gaussian model:
\begin{align}
  \src \sim \distcmpnormal{0}{\psd}. \label{eq:src}
\end{align}
By marginalizing $\src$ from Eqs.~\eqref{eq:spk-tf-mixture} and \eqref{eq:src}, the mixture signal model is obtained as:
\begin{align}
  \x@ \sim \distcmpnormal{\bm{0}}{\mathbf{Y}_{ft}}, \label{eq:lgm}
\end{align}
where $\mathbf{Y}_{ft}\triangleq \sum_{n=1}^N \msk \psd \scm$ and $\scm = \sv \sv^\adj \in \setSp^{M\times M}$ is called the spatial covariance matrix (SCM)~\cite{duongUnderDeterminedReverberantAudio2010,sawadaMultichannelExtensionsNonNegative2013} of source $n$ at frequency $f$ that characterizes the signal propagation under a certain geometric configuration. 

To facilitate the inference, SCMs are further assumed to be jointly diagonalizable (JD) by a common projection matrix $\Q \in \setC^{M\times M}$ across all the sources~\cite{sekiguchiFastMultichannelNonnegative2020,itoFastMNMFJointDiagonalization2019}:
\begin{align}
  \scm = \Q^{-1} \diag(\g@) \Q^{-\adj}, \label{eq:jd-scm}
\end{align}
where $\g@ \in \setRp^M$ are the diagonal coefficients for source $n$.
Both $\Q[] \triangleq \{ \Q \}_{f=1}^F$ and $\g* \triangleq \{\g@\}_{n,f=1}^{N,F}$ can be efficiently estimated by the iterative source steering (ISS) algorithm~\cite{scheiblerFastStableBlind2020,sekiguchiFastMultichannelNonnegative2020}.

Finally, the deep spectral modeling is applied to the source PSDs $\psd$. It introduces $D$-dimensional latent source features $\z@ \in \setR^D$ to generate the PSD $\psd$ with a deep neural network (DNN) $\dec: \setR^D \rightarrow \setRp$ as follows:
\begin{align}
  \psd = \dec(\z@), \label{eq:dsp}
\end{align}
where $\theta$ represents the model parameters of $\dec$ and the latent features $\z@$ are supposed to be spectral characteristics (e.g., pitches and envelopes) of the $n$-th source. Such a neural generative model can be trained as the decoder of a variational autoencoder (VAE)~\cite{leglaiveRecurrentVariationalAutoencoder2020,bandoNeuralFullRankSpatial2021,liFastMVAE2ImprovingAccelerating2022,kingmaAutoEncodingVariationalBayes2022} for isolated signals by assuming 
$\z@ \sim \distnormal{\bm{0}}{\eye}$.

\subsection{Inference model}
Given the observed mixture signals $\x*\triangleq \{\x@\}_{f,t=1}^{F,T}$ as input, an inference model $h_\phi$ is designed to estimate its latent source features $\z*\triangleq \{\z@\}_{n,t=1}^{N,T}$, speaker activity masks $\msk* \triangleq \{u_{nt}\}_{n,t=1}^{N,T}$, projection matrices $\Q[]$ and the diagonal coefficients $\g*$. To this end, $h_\phi$ is constructed in a hybrid manner, which contains a neural encoder modeling the posterior distribution  $q_\phi(\z*|\x*), q_\phi(\msk*|\x*)$ of $\z*$ and $\msk*$ and an ISS diagonalizer~\cite{scheiblerSurrogateSourceModel2021} computing $\Q$ and $\g@$ 
\begin{align}
  \left\{ \Q[], \g*, q_\phi(\z*|\x*), q_\phi(\msk*|\x*) \right\} \leftarrow h_\phi(\x*). \label{eq:inference}
\end{align}
where the posterior variational distribution $q_\phi(\z*|\x*)$ is assumed to follow Gaussian distribution parametrized by mean $\mu_{\phi,ntd} \in \setR$ and variance $\sigma^2_{\phi,ntd} \in \setRp$ and $ q_\phi(\msk*|\x*)$ to Bernoulli distribution parametrized by the logits $\eta_{\phi,nt} \in [0, 1]$.

Once all the parameters are inferred from Eq.~\eqref{eq:inference}, the separation is done by estimating isolated image signals $\mathbf{x}_{nft}$ using a multichannel Wiener filter~\cite{bandoNeuralFastFullRank2023,sekiguchiFastMultichannelNonnegative2020} 
$\mathbf{x}_{nft}=\msk\mathbf{Y}_{nft}\mathbf{Y}_{ft}^{-1}\x@$ and the diarization is done by thresholding the logits $\eta_{\phi,nt}$ to 0 or 1. 

\section{Heavy-tailed neural blind source separation and diarization}
We propose the heavy-tailed neural FCASA as an extension to the neural FCASA in Eq.~\eqref{eq:src} to better estimate the PSD of the isolated signals by heavy-tailed models in order to improve the diarization performance. 

\subsection{Heavy-tailed modeling for power spectrum density}
A typical variance modeling in source separation assumes the Gaussianity of the signal's time-frequency representation. However, heavy-tailed models show better robustness against impulsive noise or uncommon scenario~\cite{fontaineGeneralizedFastMultichannel2022}. 
Moreover, the real meeting recordings have a wider dynamic range of the overlap ratio~\cite{watanabeCHiME6ChallengeTackling2020} comparing to full-overlap signals in source separation task, hence such signals have a stronger heavy-tailed property. 

\begin{figure}
    \centering
    \includegraphics[width=\linewidth]{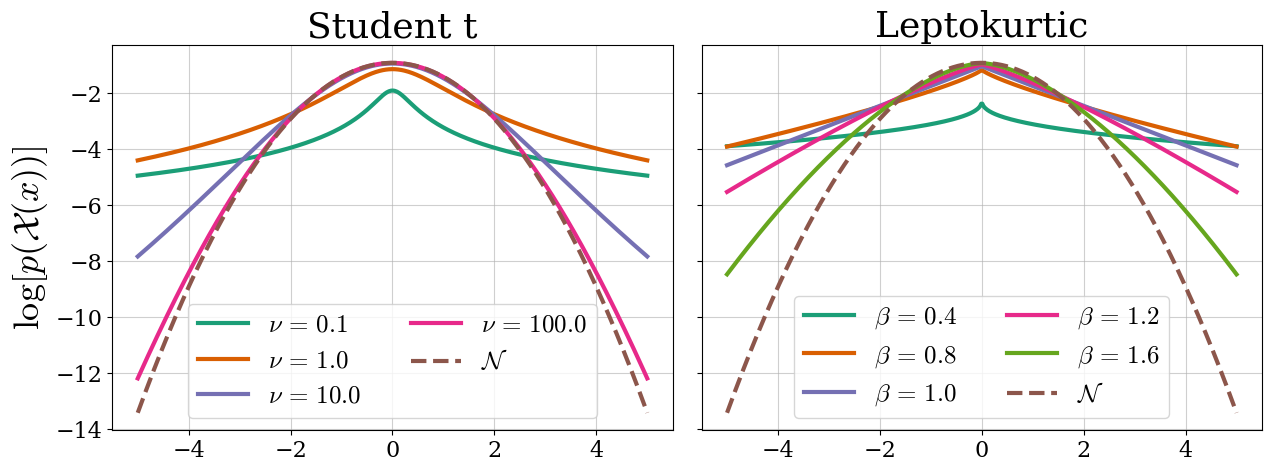}
    \caption{Standard univariate heavy-tailed models}\small{Left: Student’s $t$ distributions with degrees of freedom $\nu \in \mathbb{R}^+$. Right: Leptokurtic Generalized Gaussian distributions with shape parameters $\beta \in (0, 2)$. They tend to Gaussian when $\nu \to +\infty$ and $\beta \to 2$, respectively}
    \label{fig:heavy-tailed}
\end{figure}

A heavy-tailed extension can be simply formulated by adding a random  positive impulse variable $r_{ft}$ to the Gaussian variance, leading to a (circularly-symmetric) Gaussian Scale Mixture~(GSM) model:
\begin{align}
    \src|r_{ft} \sim \distcmpnormal{0}{r_{ft} \psd} \label{eq:hv-src}
\end{align}
where we can assume different prior distribution of $r_{ft}$. By marginalizing $r_{ft}$, we can get a specific heavy-tailed model for the mixture signal $\x@$. Particularly in this paper, we are interested in two common heavy-tailed model families, the Leptokurtic Generalized Gaussian distribution and the Student's t distribution~\cite{fontaineGeneralizedFastMultichannel2022} (Fig. \ref{fig:heavy-tailed}):
\begin{align}
\begin{aligned}
    \x@ &\sim \distcmplepto{0}{\mathbf{Y}_{ft}} &\text{(Leptokurtic)}, \\
    \x@ &\sim \distcmpstudent{0}{\mathbf{Y}_{ft}} &\text{(Student's t)},
\end{aligned}
\label{eq:hv-tf-mixture}
\end{align}
where $\beta \in (0,2), \nu \in \mathbb{R}^+$ are shape parameters for Leptokurtic Generalized Gaussian and Student's t distribution respectively.
Note that the GSM reduces to a Gaussian model hence to the vanilla neural FCASA when $r_{ft}$ is deterministic. 

\subsection{Training and inference}
We train the model parameters in Eq.~\eqref{eq:hv-tf-mixture} by maximizing its log-likelihood (LL) $\log p_{\theta}(\x*, \msk*)$ of the mixture observations $\x*$ and the speaker label $\msk*$, which is similar to~\cite{bandoNeuralFastFullRank2023, bandoNeuralBlindSource2024}. 
We use the amortized variational inference to optimize the evidence lower bound~(ELBO) of the LL \cite{kingmaAutoEncodingVariationalBayes2022, kingmaSemisupervisedLearningDeep2014}: 
\begin{align}
\begin{aligned}
&\log p_{\theta}(\x*,\msk*) 
	\geq\underbrace{\mathbb{E}_{q_{\phi}(\z*\mid\x*,\msk*)}[\log p_{\theta}(\x*\mid\msk*,\z*)]}_{\triangleq \mathcal{L}_1}\\
	&\underbrace{-\mathrm{KL}[q_{\phi}(\z*\mid\x*,\msk*)\mid\mid p_{\theta}(\z*)]}_{\triangleq \mathcal{L}_2}\underbrace{+\log p(\mathbf{U})}_{\triangleq \mathcal{L}_3},
\end{aligned}
\end{align}
where $\mathrm{KL}$ stands for the Kullback-Leiber~(KL) divergence. For simplicity, we approximate the variational family $q_\phi(\z*|\x*,\msk*)$ by the inference model $q_\phi(\z*|\x*)$.

For the first term, we use the one-point Monte-Carlo estimation with one sample $\z*$ sampled from $q(\z*|\x*)$ and $\msk*$ directly from the ground-truth label as teacher-forcing. According to \cite{fontaineGeneralizedFastMultichannel2022}, the conditional likelihood is:
\begin{align}
    \mathcal{L}_1=\sum_{f,t=1}^{F,T} \tilde{\mathcal{L}}_1 + T\sum_{f=1}^{F} \log |\Q\Q^\adj|,
\end{align}
where $\tilde{\mathcal{L}}_1$ is (up to an additive constant)
\begin{align}
    -\left(\sum_{m=1}^M \frac{|\xt|^2}{\yt[:ftm]}\right)^{\frac{\beta}{2}}-\sum_{m=1}^M \log  \yt[:ftm],
\end{align} for Leptokurtic Generalized Gaussian distribution and
\begin{align}
-\left(\frac{\nu}{2}+M \right)\log \left( 1+\frac{2}{\nu}\sum_{m=1}^M \frac{|\xt|^2}{\yt[:ftm]}\right)-\sum_{m=1}^M \log  \yt[:ftm],
\end{align}
for Student's t distribution,
where $\xt,\yt$ are the $m$-th entry of $\xt@\triangleq \Q\x@,\yt@ \triangleq \msk \dec(\z@) \g@$ and $\yt[:ftm] \triangleq \sum_{n=1}^{N} \yt$.

The second term is a KL-divergence between two Gaussians and can be computed in closed form~\cite{kingmaAutoEncodingVariationalBayes2022}. For the third term, as we don't assume a prior model $p(\msk*)$ for the speaker activity mask $\msk*$ \cite{maitiEENDSSJointEndtoEnd2022,bandoNeuralBlindSource2024}, instead, we directly maximize the log-posterior of surrogate model, $\log q_\phi(\msk*|\x*)$, which corresponds to minimize the KL-divergence between the empirical posterior distribution, $p_{data}(\msk*|\x*)$ and the surrogate posterior, $q_\phi(\msk*|\x*)$, which is equivalent to minimizing the binary cross entropy (BCE) between the ground truth label, $u_{nt}$, and the logit $\eta_{\phi,nt}$, i.e.
\begin{align}
    \mathcal{L}_3 
    \approx \log q_\phi(\msk*|\x*) = -\sum_{n,t=1}^{N,T}\text{BCE}[\msk|\eta_{\phi, nt}].
\end{align}
Note that this shares the same insights as the extended objective function in the semi-supervised VAE \cite{kingmaSemisupervisedLearningDeep2014}, where we want to train the inference model for $\msk*$ directly from the labelled data as well.

Meanwhile, the permutation invariant
training (PIT) is applied to solve the speaker permutation ambiguity~\cite{maitiEENDSSJointEndtoEnd2022}.
Combining them altogether, we deduce the final form of our maximization objective:
\begin{align}
    \mathcal{L} = \frac{1}{TF} \left(\mathcal{L}_1+\delta\mathcal{L}_2\right) + \gamma\frac{1}{TN} \mathcal{L}_3,
    \label{eq:objective}
\end{align}
where $\delta$ and $\gamma$ are scaling hyperparameters. It turns out that the heavy-tailed neural FCASA can fully inherit the neural FCASA training and inference procedure by just replacing the original objective to Eq.~\eqref{eq:objective}, leading to a straightforward extension.

\section{Experiments}
\label{sec:results}
To fairly compare with the neural FCASA as our baseline model, we follow the same experimental settings as in~\cite{bandoNeuralBlindSource2024} \footnote{Codes available at \url{https://github.com/alephpi/neural-fcasa}.}.
\subsection{Dataset description}
\label{sec:dataset}
We evaluate the proposed method on three public corpora that provide a multi-channel far-field subset: 
\begin{enumerate}
    \item AMI corpus~\cite{carlettaAMIMeetingCorpus2005}.
    This dataset contains 100 hours of 16 kHz English meeting recordings with 3 to 5 participants. Audio was captured using an 8-mic circular array with a radius of 10 cm placed on the table. Our experiments adopt the official split: \verb|training| (80.7h) for training, \verb|development| (9.7h) for validation and \verb|evaluation| (9.1h) for evaluation.
    \item AliMeeting corpus~\cite{yu2022alimeeting}. This dataset contains 118.75 hours of 16 kHz Mandarin meeting recordings with 2 to 4 participants. Audio was captured using an 8-mic circular array with a radius of 5 cm placed on the table. Our experiments adopt the official split: \verb|training| (104.75h) for training, \verb|evaluation| (4h) for validation and \verb|test| (10h) for evaluation.\footnote{Note that the dataset split naming convention of AliMeeting is different from the one of AMI or CHiME-6.} Besides, we use the aligned RTTM transcription from~\cite{horiguchi_asru2025} as the ground truth labels instead of the original one.
    \item CHiME-6 corpus~\cite{watanabe20b_chime} contains 50 hours of 16 kHz English dinner party conversation recordings with 4 participants. Audio was captured using 6 distant Microsoft Kinect devices. Each device has a linear array of 4 sample-synchronized microphones. The participants are moving around different rooms (kitchen, dining room and living room) in the house. Our experiments adopt the official split: \verb|training| (40.5h) for training, \verb|development| (4.5h) for validation and \verb|evaluation| (5h) for evaluation. For a fair comparison with AMI and AliMeeting, we only use the recordings from device 1 and 2 and merge them as 8 channel recordings.
\end{enumerate}
\subsection{Configurations}
\label{sec:config}
The network architecture of the heavy-tailed neural FCASA follows that of the vanilla neural FCASA and we refer the reader to~\cite{bandoNeuralBlindSource2024} for more details.
%
All the signals are dereverberated in advance using weighted prediction error (WPE)~\cite{yoshiokaGeneralizationMultichannelLinear2012}. The spectrograms are generated via STFT with a window size of 512 and a hop size of 160.
The source number $N=6$ containing 5 speaker channels 
and a noise channel. 
The latent dimension $D=10$ for the noise channel and $D=64$ for the speaker channels to avoid channel modeling ambiguity~\cite{bando22_interspeech}. Speaker activations $\msk$ are derived from the oracle diarization labels, while the noise channel is always set to active to prevent confusion with silence. $\delta,\gamma$ in Eq.~\eqref{eq:objective} are both set to 1.0.
During training, the audio recordings are segmented into 20s clips, from which 10s continuous crops are randomly sampled and fed into the training pipeline with a batch size of 128. 
The model is trained for 200 epochs using the AdamW optimizer with a learning rate of 1e-4 and a weight decay of 1e-5. 
For evaluation, the audio recordings are segmented into 10s clips and fed into the inference pipeline. The speaker activity predictions $\eta_{\phi,nt}$ are smoothed by a median filter of 11 frames for stability and then cast to 0 or 1 using a threshold of 0.5.

We use Pyannote~\cite{BredinPyannote2023} to report the diarization error rate (DER) and its three components: miss, false alarm (FA) and speaker confusion (Conf.), as well as the Jaccard Error Rate (JER).  \footnote{Note that we do not evaluate the objective source separation metrics since all of these corpora consist of real conversations without isolated ground truth.}
To evaluate the diarization performance under different standards, we adopt four different evaluation setups. Three of them follow the convention proposed in~\cite{landiniBayesianHMMClustering2022}, where Forgiving applies a 0.25s collar without overlap, Fair applies a 0.25s collar within overlap and Full applies no collar within overlap. In addition, we adopt the Overlap setup, which applies no collar but evaluates only on the overlap, since the overlapping part of speech is a major challenge for diarization systems.  

We first train all our models on the AMI corpus training set. We have trained five models in the Leptokurtic Generalized Gaussian family and four in the Student's t family, with shape parameters $\beta\in\{0.4, 0.8, 1.0, 1.2, 1.6\}$ and $\nu\in\{0.1, 1, 10, 100\}$ respectively. 
We also retrained a baseline Gaussian model proposed in the neural FCASA.
Note that the Gaussian model can be regarded as Leptokurtic Generalized Gaussian with $\beta=2$ or Student's t with $\nu \to \infty$. For each model setting, we track the diarization loss $\mathcal{L}_3$ on the validation set and find the best checkpoints and evaluate all the metrics on the evaluation set. 

Once found the optimal parameter settings on the AMI corpus, we launch the model training and evaluate their performances on AliMeeting corpus and CHiME-6 corpus to verify the generalization ability of our method. We also perform cross-tests for models trained on one corpus and test them on another corpus.
\begin{table}[H]
\caption{Diarization metrics of the trained models on AMI evaluation set. All metrics are reported in percentage with the best score in \textbf{bold} and the second-best score \uline{underlined}. We also list the single channel diarization baseline Pyannote 3.1 \cite{BredinPyannote2023} as a reference. Note that the evaluation of our method is performed chunk-wisely (10 seconds), which is not directly comparable to the Pyannote baselines.}
\resizebox{\columnwidth}{!}{
\begin{tabular}{@{}cclccccc@{}}
\toprule
Model &
  \begin{tabular}[c]{@{}c@{}}Shape \\ param.\end{tabular} &
  \multicolumn{1}{c}{Setup} &
  Miss ($\downarrow$) &
  FA ($\downarrow$) &
  Conf. ($\downarrow$) &
  DER ($\downarrow$) &
  JER ($\downarrow$) \\ \midrule
\multirow{4}{*}{\begin{tabular}[c]{@{}c@{}}Gaussian\\ (baseline)\end{tabular}} &
  \multirow{4}{*}{-} &
  Forgiving &
  5.59 &
  8.16 &
  0.73 &
  14.48 &
  13.50 \\
                              &                      & Fair      & 8.61           & 6.33          & 0.79          & 15.73          & 23.16          \\
                              &                      & Full      & 10.77          & 6.81          & 1.16          & 18.73          & 27.65          \\
                              &                      & Overlap   & 19.40          & 3.80          & 0.90          & 24.11          & 26.53          \\ \midrule
\multirow{20}{*}{Leptokurtic} & \multirow{4}{*}{0.4} & Forgiving & 7.98           & 6.23          & 0.86          & 15.07          & 15.81          \\
                              &                      & Fair      & 11.52          & 4.97          & 0.92          & 17.41          & 27.33          \\
                              &                      & Full      & 13.85          & 5.42          & 1.28          & 20.54          & 31.52          \\
                              &                      & Overlap   & 24.07          & 3.05          & 0.90          & 28.02          & 31.35          \\ \cmidrule(l){2-8} 
                              & \multirow{4}{*}{0.8} & Forgiving & 6.14           & 7.84          & 0.83          & 14.81          & 13.98          \\
                              &                      & Fair      & 9.35           & 6.29          & 0.96          & 16.61          & 24.36          \\
                              &                      & Full      & 11.63          & 6.82          & 1.30          & 19.75          & 28.70          \\
                              &                      & Overlap   & 20.79          & 4.20          & 0.96          & 25.95          & 28.10          \\ \cmidrule(l){2-8} 
                              & \multirow{4}{*}{1.0} & Forgiving & 5.45           & 7.36          & 0.79          & 13.60          & 13.01          \\
                              &                      & Fair      & 8.84           & 5.76          & 0.80          & 15.41          & 22.66          \\
                              &                      & Full      & 11.08          & 6.26          & 1.13          & 18.47          & 27.38          \\
                              &                      & Overlap   & 20.13          & 3.52          & 0.80          & 24.45          & 27.02          \\ \cmidrule(l){2-8} 
                              & \multirow{4}{*}{1.2} & Forgiving & 5.13           & 9.12          & 0.86          & 15.11          & 13.90          \\
                              &                      & Fair      & 8.30           & 6.80          & 0.95          & 16.06          & 23.59          \\
                              &                      & Full      & 10.38          & 7.58          & 1.34          & 19.30          & 28.17          \\
                              &                      & Overlap   & 18.98          & 4.11          & 1.03          & 24.12          & 26.78          \\ \cmidrule(l){2-8} 
                              & \multirow{4}{*}{1.6} & Forgiving & 6.04           & 6.65          & 0.71          & 13.39          & 13.32          \\
                              &                      & Fair      & 9.55           & 5.08          & 0.75          & 15.39          & 23.33          \\
                              &                      & Full      & 11.82          & 5.65          & 1.06          & 18.52          & 27.92          \\
                              &                      & Overlap   & 21.09          & 3.10          & 0.74          & 24.93          & 27.82          \\ \midrule
\multirow{16}{*}{Student's t} &
  \multirow{4}{*}{0.1} &
  Forgiving &
  {\ul 5.04} &
  \textbf{5.59} &
  \textbf{0.61} &
  \textbf{11.24} &
  \textbf{10.78} \\
                              &                      & Fair      & {\ul 7.49}     & \textbf{4.42} & \textbf{0.62} & \textbf{12.53} & \textbf{18.38} \\
                              &                      & Full      & {\ul 9.23}     & {\ul 5.86}    & \textbf{0.92} & \textbf{16.01} & \textbf{23.45} \\
                              &                      & Overlap   & {\ul 16.18}    & \textbf{3.18} & \textbf{0.67} & {\ul 20.03}    & {\ul 21.99}    \\ \cmidrule(l){2-8} 
                              & \multirow{4}{*}{1}   & Forgiving & \textbf{4.30}  & {\ul 6.44}    & {\ul 0.67}    & {\ul 11.41}    & {\ul 11.04}    \\
                              &                      & Fair      & \textbf{6.92}  & {\ul 5.02}    & {\ul 0.69}    & {\ul 12.62}    & {\ul 18.80}    \\
                              &                      & Full      & \textbf{8.55}  & 6.52          & {\ul 0.98}    & {\ul 16.05}    & {\ul 23.82}    \\
                              &                      & Overlap   & \textbf{15.57} & {\ul 3.44}    & {\ul 0.70}    & \textbf{19.71} & \textbf{21.92} \\ \cmidrule(l){2-8} 
                              & \multirow{4}{*}{10}  & Forgiving & 5.45           & 7.58          & 0.78          & 13.80          & 12.17          \\
                              &                      & Fair      & 8.55           & 6.17          & 0.80          & 15.52          & 20.98          \\
                              &                      & Full      & 10.93          & 5.99          & 1.07          & 17.99          & 25.95          \\
                              &                      & Overlap   & 19.76          & 3.71          & 0.74          & 24.21          & 26.01          \\ \cmidrule(l){2-8} 
                              & \multirow{4}{*}{100} & Forgiving & 5.57           & 6.78          & 0.72          & 13.08          & 12.48          \\
                              &                      & Fair      & 8.93           & 5.39          & 0.80          & 15.12          & 22.09          \\
                              &                      & Full      & 11.25          & {\ul 5.96}    & 1.09          & 18.30          & 26.82          \\
                              &                      & Overlap   & 20.12          & 3.55          & 0.78          & 24.44          & 26.72          \\ \bottomrule
\multicolumn{2}{l}{{\begin{tabular}[c]{@{}l@{}}\grey{Pyannote 3.1}\\ \grey{IDM}\end{tabular}}} &
  \grey{Full} &
  \grey{9.5} &
  \grey{3.6} &
  \grey{5.7} &
  \grey{18.8} &
  \grey{-} \\
\multicolumn{2}{l}{\begin{tabular}[c]{@{}l@{}}\grey{Pyannote 3.1}\\ \grey{SDM}\end{tabular}} &
  \grey{Full} &
  \grey{11.2} &
  \grey{3.8} &
  \grey{7.5} &
  \grey{22.4} &
  \grey{-} \\ \hline
\end{tabular}
}

\label{tab:results_ami}
\end{table}
\subsection{Results and discussions}
The evaluation results on AMI corpus are illustrated in Table~\ref{tab:results_ami}.
We first observe that for each model, the error rates become larger when the evaluation setup becomes stricter on overlaps. We then observe some heavy-tailed models have consistent improvements against the baseline in error rates across different setups. For example, we have better results from Leptokurtic Generalized Gaussian models with $\beta=1.0, 1.6$, whereas Student's t models outperform the baseline by a large margin (2-3\% in absolute and 10-15\% in relative) with $\nu=\{0.1,1\}$. This validates the effectiveness of the proposed heavy-tailed neural FCASA method and also demonstrates the choice of the shape parameters substantially affects model performance. Note that for both heavy-tailed families, a smaller parameter implies a heavier tail. In particular, when $\nu=1$, the Student's t corresponds to the Cauchy distribution, which preserves the law under linear combination, making it naturally compatible with the linear model assumption in Eq.~\eqref{eq:spk-tf-mixture}. 
Larger shape values ($\nu=100, \beta=1.6$) converge towards Gaussian model leading to worse performance.
Also, there are high potentials that we could find a better performing shape parameter when $\beta\in [1.0,1.6]$ or $\nu \in [0.1,1]$.

\begin{table}[h]
\caption{Diarization metrics of the trained models on AliMeeting test set. All metrics are reported in percentage with the best score in bold and the second-best score underlined. We also list the single channel diarization baseline Pyannote 3.1 \cite{BredinPyannote2023} as a reference. Note that the evaluation of our method is performed chunk-wisely (10 seconds), which is not directly comparable to the Pyannote baseline.}
\resizebox{\columnwidth}{!}{
\begin{tabular}{cclccccc}
\hline
Model &
  \begin{tabular}[c]{@{}c@{}}Shape \\ param.\end{tabular} &
  \multicolumn{1}{c}{Setup} &
  Miss ($\downarrow$) &
  FA ($\downarrow$) &
  Conf. ($\downarrow$) &
  DER ($\downarrow$) &
  JER ($\downarrow$) \\ \hline
\multirow{4}{*}{\begin{tabular}[c]{@{}c@{}}Gaussian\\ (baseline)\end{tabular}} &
  \multirow{4}{*}{-} &
  Forgiving &
  2.52 &
  2.53 &
  0.37 &
  5.42 &
  7.04 \\
 &  & Fair    & 3.46           & 1.90          & 0.40          & 5.75           & 10.29          \\
 &  & Full    & 7.25           & {\ul 4.52}    & 0.70          & 12.47          & 19.89          \\
 &  & Overlap & 14.98          & {\ul 2.00}    & 0.47          & 17.46          & {\ul 20.25}    \\ \hline
\multirow{8}{*}{Student's t} &
  \multirow{4}{*}{0.1} &
  Forgiving &
  {\ul 2.28} &
  {\ul 2.35} &
  \textbf{0.22} &
  {\ul 4.85} &
  {\ul 5.97} \\
 &  & Fair    & {\ul 2.93}     & {\ul 1.76}    & \textbf{0.22} & {\ul 4.91}     & \textbf{8.46}  \\
 &  & Full    & \textbf{6.38}  & 4.60          & \textbf{0.49} & {\ul 11.47}    & \textbf{18.11} \\
 &  & Overlap & \textbf{13.42} & 2.01          & {\ul 0.35}    & \textbf{15.78} & \textbf{18.28} \\ \cline{2-8} 
 &
  \multirow{4}{*}{1} &
  Forgiving &
  \textbf{1.87} &
  \textbf{2.16} &
  {\ul 0.28} &
  \textbf{4.32} &
  \textbf{5.88} \\
 &  & Fair    & \textbf{2.62}  & \textbf{1.59} & {\ul 0.28}    & \textbf{4.49}  & {\ul 8.47}     \\
 &  & Full    & {\ul 6.41}     & \textbf{4.09} & {\ul 0.52}    & \textbf{11.03} & {\ul 18.18}    \\
 &  & Overlap & {\ul 14.17}    & \textbf{1.76} & \textbf{0.33} & {\ul 16.26}    & {\ul 18.88}    \\ \hline

\multicolumn{2}{l}{\begin{tabular}[c]{@{}l@{}}\grey{Pyannote 3.1}\\ \grey{channel 1}\end{tabular}} &
  \grey{Full} &
  \grey{10.0} &
  \grey{4.4} &
  \grey{10.0} &
  \grey{24.4} &
  \grey{-} \\ \hline
\end{tabular}}

\label{tab:results_ali}
\end{table}

\begin{table}[h]
\caption{Diarization metrics of the trained models on CHiME-6 evaluation set. All metrics are reported in percentage with the best score in bold and the second-best score underlined. We also list the single channel diarization official baseline from CHiME-6 \cite{watanabeCHiME6ChallengeTackling2020} as a reference. Note that the evaluation of our method is performed chunk-wisely (10 seconds), which is not directly comparable to the CHiME-6  baseline.}
\resizebox{\columnwidth}{!}{
\begin{tabular}{cclccccc}
\hline
Model &
  \begin{tabular}[c]{@{}c@{}}Shape \\ param.\end{tabular} &
  \multicolumn{1}{c}{Setup} &
  Miss ($\downarrow$) &
  FA ($\downarrow$) &
  Conf. ($\downarrow$) &
  DER ($\downarrow$) &
  JER ($\downarrow$) \\ \hline
\multirow{4}{*}{\begin{tabular}[c]{@{}c@{}}Gaussian\\ (baseline)\end{tabular}} &
  \multirow{4}{*}{-} &
  Forgiving &
  {\ul 10.73} &
  61.51 &
  6.96 &
  79.20 &
  42.08 \\
 &                    & Fair      & \textbf{14.72} & 38.46          & 7.46          & 60.65                & 49.27          \\
 &                    & Full      & \textbf{17.48} & 35.61          & 8.64          & 61.74                & 52.89          \\
 &                    & Overlap   & {\ul 29.84}    & 6.74           & 6.56          & {\ul 43.14}          & {\ul 51.68}    \\ \hline
\multirow{8}{*}{Student's t} &
  \multirow{4}{*}{0.1} &
  Forgiving &
  \textbf{9.16} &
  {\ul 30.47} &
  \textbf{5.75} &
  {\ul 45.37} &
  \textbf{32.40} \\
 &                    & Fair      & {\ul 14.83}    & {\ul 19.30}    & \textbf{6.22} & {\ul 40.35}          & \textbf{41.67} \\
 &                    & Full      & {\ul 17.85}    & {\ul 29.01}    & \textbf{7.23} & {\ul 54.10}          & \textbf{46.61} \\
 &                    & Overlap   & \textbf{27.15} & {\ul 10.95}    & {\ul 6.14}    & 44.25                & \textbf{46.60} \\ \cline{2-8} 
 & \multirow{4}{*}{1} & Forgiving & 15.89          & \textbf{19.20} & {\ul 6.58}    & \textbf{41.67}       & {\ul 38.41}    \\
 &                    & Fair      & 20.52          & \textbf{12.54} & {\ul 6.80}    & \textbf{39.86}       & {\ul 46.97}    \\
 &                    & Full      & 23.72          & \textbf{15.54} & {\ul 7.57}    & \textbf{46.84}       & {\ul 51.44}    \\
 &                    & Overlap   & 32.61          & \textbf{3.66}  & \textbf{6.08} &  \textbf{42.35} & 51.89          \\ \hline

\multicolumn{2}{l}{\begin{tabular}[c]{@{}l@{}}\grey{CHiME-6 official baseline}\\ \grey{Single channel}\end{tabular}} &
  \grey{Full} &
  \grey{-} &
  \grey{-} &
  \grey{-} &
  \grey{62.0} &
  \grey{71.4} \\ \hline
\end{tabular}}
\label{tab:results_chime6}
\end{table}

The performances of models trained respectively on AliMeeting and CHiME-6 training sets with optimal parameter settings found on AMI corpus (i.e. Student's t distribution with $\nu\in\{0.1,1\}$) are reported in Table~\ref{tab:results_ali} and Table~\ref{tab:results_chime6}. We observe consistent improvements against the baseline. For AliMeeting corpus, we observe 1\%-2\% in absolute (10\%-20\% in relative) improvements compared to the baseline. For CHiME-6 corpus, we even observe up to 20\% 
in absolute (35\% in relative) improvements compared to the baseline, due to the difficulty of CHiME-6 corpus and the relatively poor performance of the baseline. Therefore, this verifies the generalization ability of our proposed method and especially justifies the necessity of the heavy-tailed models for difficult acoustic environments.

The cross-tests results are reported in Table~\ref{tab:results_cross_tests}. We observe that for train-test pairs such as AMI-AliMeeting and AliMeeting-AMI, there are minor but consistent improvements over the baseline. However, the performances are still far worse than the in-domain results. This can be justified by the similarity of the meeting scenario and the microphone array configuration between AMI and Alimeeting and the major mismatch is the working language (English vs. Mandarin). On the other hand, cross-tests involving CHiME-6 show less consistency, we think this is due to the large gap between the scenario (meeting vs. party, steady vs. moving), the microphone array configuration (circular vs. linear) and language.  

\begin{table}[t]
\caption{Diarization metrics of the cross-tests. All metrics are reported  percentage. We only select the DER of the Full setup for simplicity.}
\begin{tabular}{cclccc}
\hline
\multirow{2}{*}{Model} &
  \multirow{2}{*}{\begin{tabular}[c]{@{}c@{}}Shape \\ param.\end{tabular}} &
  \multicolumn{1}{r}{test} &
  \multirow{2}{*}{AMI} &
  \multirow{2}{*}{AliMeeting} &
  \multirow{2}{*}{CHiME-6} \\
                             &                      & train      &       &       &       \\ \hline
\multirow{3}{*}{\begin{tabular}[c]{@{}c@{}}Gaussian\\ (baseline)\end{tabular}} &
  \multirow{3}{*}{-} &
  AMI &
  18.73 &
  41.99 &
  80.53 \\
                             &                      & AliMeeting & 44.11  & 12.47 & 70.71 \\
                             &                      & CHiME-6    & 47.06 & 38.17 & 61.74 \\ \hline
\multirow{6}{*}{Student's t} & \multirow{3}{*}{0.1} & AMI        & 16.01 & 39.35 & 64.25 \\
                             &                      & AliMeeting & 43.28 & 11.47 & 74.41 \\
                             &                      & CHiME-6    & 75.45 & 43.15 & 54.10 \\ \cline{2-6} 
                             & \multirow{3}{*}{1}   & AMI        & 16.05 & 38.48 & 80.96 \\
                             &                      & AliMeeting & 43.33 & 11.03 & 65.52 \\
                             &                      & CHiME-6    & 40.21 & 35.94 & 46.84 \\ \hline
\end{tabular}
\label{tab:results_cross_tests}
\end{table}

\subsection{Limitations}
We notice that for some parameter configurations of Leptokurtic Generalized Gaussian distributions, there is a diarization performance degradation compared to the baseline, e.g. for $\beta\in\{0.4,0.8,1.2\}$. This suggests the potential sensitivity of our method to hyperparameter tuning. We will leave the investigation and improvements as future work such as direct estimation of those parameters. 


\section{Conclusion and future work}
We propose a heavy-tailed extension to the neural FCASA system which jointly performs speaker separation and diarization. The proposed method can be formulated as a Gaussian scale mixture model with a random impulse variable to the variance, resulting in a fairly simple objective modification of the original baseline. Through extensive experiments, we find various parameter settings that outperform the baseline by a large margin. In future work, we would like to explore the extension to more heavy-tailed model families and their performance on speaker separation. Alternative directions include finding a better shape parameter tuning strategy, incorporating the speaker activity model into the heavy-tailed modeling \cite{mao2026neuralmultichanneldistantspeaker} and incorporating the geometrical knowledge of microphone array~\cite{sumuraJointAudioSource2024}. 


\section{Acknowledgement}
We acknowledge the use of Claude (https://claude.ai/) to polish this paper. We pasted the draft of the abstract, the first three paragraphs of Section \ref{sec:intro}, Section \ref{sec:dataset} and Section \ref{sec:config} with the instruction ``Please rephrase the text in an academic manner''. The outputs were then modified further to better represent our own writing.

We thank Dr. Yoshiaki Bando for insightful discussions. 
This project is funded by ANR Project SAROUMANE (ANR-22-CE23-0011) and granted access to the HPC resources of IDRIS under the allocation 20191014876 attribution made by GENCI.
\section{Reference}
\printbibliography[heading=none]

@string{icassp = "Proc. ICASSP"}

@string{interspeech = "Proc. Interspeech"}

@string{iwaenc = "Proc. IWAENC"}

@string{eusipco = "Proc. EUSIPCO"}

@string{asru = "Proc. ASRU"}

@string{ieee-taslp = "IEEE Trans. Audio, Speech, Lang. Process."}

@string{ieee-acm-taslp = "IEEE/ACM Trans. Audio, Speech, Lang. Process."}

@string{ieee-spl = "IEEE Signal Process. Lett."}

@string{iclr = "Proc. ICLR"}

@string{csl = "Comput. Speech Lang."}

@string{slt = "Proc. SLT"}

@inproceedings{bandoNeuralBlindSource2024,
  title     = {{Neural Blind Source Separation and Diarization for Distant Speech Recognition}},
  author    = {Yoshiaki Bando and Tomohiko Nakamura and Shinji Watanabe},
  year      = {2024},
  booktitle = interspeech,
  pages     = {722--726},
  doi       = {10.21437/Interspeech.2024-1137},
  issn      = {2958-1796},
}

@inproceedings{bandoNeuralFastFullRank2023,
  title = {Neural {{Fast Full-Rank Spatial Covariance Analysis}} for {{Blind Source Separation}}},
  booktitle = eusipco,
  author = {Bando, Yoshiaki and Masuyama, Yoshiki and Nugraha, Aditya Arie and Yoshii, Kazuyoshi},
  year = {2023},
  month = sep,
  pages = {51--55},
  publisher = {IEEE},
  doi = {10.23919/EUSIPCO58844.2023.10289974},
  isbn = {978-94-645936-0-0}
}

@article{bandoNeuralFullRankSpatial2021,
  title = {Neural {{Full-Rank Spatial Covariance Analysis}} for {{Blind Source Separation}}},
  author = {Bando, Yoshiaki and Sekiguchi, Kouhei and Masuyama, Yoshiki and Nugraha, Aditya Arie and Fontaine, Mathieu and Yoshii, Kazuyoshi},
  year = {2021},
  journal = ieee-spl,
  volume = {28},
  pages = {1670--1674},
  issn = {1558-2361},
  doi = {10.1109/LSP.2021.3101699}
}

@article{duongUnderDeterminedReverberantAudio2010,
  title = {Under-{{Determined Reverberant Audio Source Separation Using}} a {{Full-Rank Spatial Covariance Model}}},
  author = {Duong, Ngoc Q K and Vincent, Emmanuel and Gribonval, R{\'e}mi},
  year = {2010},
  month = sep,
  journal = ieee-taslp,
  volume = {18},
  number = {7},
  pages = {1830--1840},
  issn = {1558-7916, 1558-7924},
  doi = {10.1109/TASL.2010.2050716}
}

@article{fontaineGeneralizedFastMultichannel2022,
  title = {{Generalized Fast Multichannel Nonnegative Matrix Factorization Based on Gaussian Scale Mixtures for Blind Source Separation}},
  author = {Fontaine, Mathieu and Sekiguchi, Kouhei and Nugraha, Aditya and Bando, Yoshiaki and Yoshii, Kazuyoshi},
  year = {2022},
  journal = ieee-acm-taslp,
  volume = {30},
  primaryclass = {cs},
  pages = {1734--1748},
  issn = {2329-9290, 2329-9304},
  doi = {10.1109/TASLP.2022.3172631},
}

@article{sawadaMultichannelExtensionsNonNegative2013,
  title = {Multichannel {{Extensions}} of {{Non-Negative Matrix Factorization With Complex-Valued Data}}},
  author = {Sawada, Hiroshi and Kameoka, Hirokazu and Araki, Shoko and Ueda, Naonori},
  year = {2013},
  month = may,
  journal = ieee-taslp,
  volume = {21},
  number = {5},
  pages = {971--982},
  issn = {1558-7924},
  doi = {10.1109/TASL.2013.2239990}
}

@inproceedings{scheiblerFastStableBlind2020,
  title = {Fast and {{Stable Blind Source Separation}} with {{Rank-1 Updates}}},
  booktitle = icassp,
  author = {Scheibler, Robin and Ono, Nobutaka},
  year = {2020},
  month = may,
  pages = {236--240},
  issn = {2379-190X},
  doi = {10.1109/ICASSP40776.2020.9053556}
}

@article{sekiguchiFastMultichannelNonnegative2020,
  title = {Fast {{Multichannel Nonnegative Matrix Factorization With Directivity-Aware Jointly-Diagonalizable Spatial Covariance Matrices}} for {{Blind Source Separation}}},
  author = {Sekiguchi, Kouhei and Bando, Yoshiaki and Nugraha, Aditya Arie and Yoshii, Kazuyoshi and Kawahara, Tatsuya},
  year = {2020},
  journal = ieee-acm-taslp,
  volume = {28},
  pages = {2610--2625},
  issn = {2329-9304},
  doi = {10.1109/TASLP.2020.3019181}
}

@incollection{vincentProbabilisticModelingParadigms2011,
  title = {Probabilistic Modeling Paradigms for Audio Source Separation},
  booktitle = {Machine {{Audition}}: {{Principles}}, {{Algorithms}} and {{Systems}}},
  author = {Vincent, Emmanuel},
  year = {2011},
  doi = {10.4018/978-1-61520-919-4}
}

@article{itoFastMNMFJointDiagonalization2019,
  title = {{{FastMNMF}}: {{Joint Diagonalization Based Accelerated Algorithms}} for {{Multichannel Nonnegative Matrix Factorization}}},
  shorttitle = {{{FastMNMF}}},
  author = {Ito, Nobutaka and Nakatani, Tomohiro},
  year = {2019},
  month = may,
  journal = icassp,
  pages = {371--375},
  publisher = {IEEE},
  address = {Brighton, United Kingdom},
  doi = {10.1109/ICASSP.2019.8682291},
  isbn = {9781479981311}
}

@article{leglaiveRecurrentVariationalAutoencoder2020,
  title = {A {{Recurrent Variational Autoencoder}} for {{Speech Enhancement}}},
  author = {Leglaive, Simon and {Alameda-Pineda}, Xavier and Girin, Laurent and Horaud, Radu},
  year = {2020},
  month = may,
  journal = icassp,
  pages = {371--375},
  publisher = {IEEE},
  address = {Barcelona, Spain},
  doi = {10.1109/ICASSP40776.2020.9053164},
  isbn = {9781509066315}
}

@inproceedings{kingmaAutoEncodingVariationalBayes2022,
  title = {Auto-{{Encoding Variational Bayes}}},
  author = {Kingma, Diederik P. and Welling, Max},
  booktitle = iclr,
  year = 2014,
  month = apr,
}

@article{liFastMVAE2ImprovingAccelerating2022,
  title = {{{FastMVAE2}}: {{On}} Improving and Accelerating the Fast Variational Autoencoder-Based Source Separation Algorithm for Determined Mixtures},
  author = {Li, Li and Kameoka, Hirokazu and Makino, Shoji},
  year = {2022},
  month = oct,
  journal = ieee-acm-taslp,
  volume = {31},
  pages = {96--110},
  publisher = {IEEE Press},
  issn = {2329-9290},
  doi = {10.1109/TASLP.2022.3214763},
  issue_date = {2023}
}

@inproceedings{watanabeCHiME6ChallengeTackling2020,
  title = {{{CHiME-6 Challenge}}: {{Tackling Multispeaker Speech Recognition}} for {{Unsegmented Recordings}}},
  booktitle = {Proc. CHiME},
  author = {Watanabe, Shinji and Mandel, Michael and Barker, Jon and Vincent, Emmanuel and Arora, Ashish and Chang, Xuankai and Khudanpur, Sanjeev and Manohar, Vimal and Povey, Daniel and Raj, Desh and Snyder, David and Subramanian, Aswin Shanmugam and Trmal, Jan and Yair, Bar Ben and Boeddeker, Christoph and Ni, Zhaoheng and Fujita, Yusuke and Horiguchi, Shota and Kanda, Naoyuki and Yoshioka, Takuya and Ryant, Neville},
  year = {2020},
  pages = {1--7},
  doi = {10.21437/CHiME.2020-1}
}

@inproceedings{maitiEENDSSJointEndtoEnd2022,
title = "{EEND-SS: Joint End-to-End Neural Speaker Diarization and Speech Separation for Flexible Number of Speakers}",
author = "Soumi Maiti and Yushi Ueda and Shinji Watanabe and Chunlei Zhang and Meng Yu and Zhang, \{Shi Xiong\} and Yong Xu",
year = "2023",
pages = "480--487",
booktitle = slt,
}

@inproceedings{bando22_interspeech,
  title     = {{Weakly-Supervised Neural Full-Rank Spatial Covariance Analysis for a Front-End System of Distant Speech Recognition}},
  author    = {Yoshiaki Bando and Takahiro Aizawa and Katsutoshi Itoyama and Kazuhiro Nakadai},
  year      = {2022},
  booktitle = interspeech,
  pages     = {{3824--3828}},
  doi       = {{10.21437/Interspeech.2022-11077}},
  issn      = {{2958-1796}},
}

@inproceedings{carlettaAMIMeetingCorpus2005,
  title = {The {{AMI}} Meeting Corpus: A Pre-Announcement},
  booktitle = {Proceedings of the Second International Conference on Machine Learning for Multimodal Interaction},
  author = {Carletta, Jean and Ashby, Simone and Bourban, Sebastien and Flynn, Mike and Guillemot, Mael and Hain, Thomas and Kadlec, Jaroslav and Karaiskos, Vasilis and Kraaij, Wessel and Kronenthal, Melissa and Lathoud, Guillaume and Lincoln, Mike and Lisowska, Agnes and McCowan, Iain and Post, Wilfried and Reidsma, Dennis and Wellner, Pierre},
  year = {2005},
  series = {{{MLMI}}'05},
  pages = {28--39},
  publisher = {Springer-Verlag},
  address = {Berlin, Heidelberg},
  doi = {10.1007/11677482_3},
  isbn = {3-540-32549-2}
}

@inproceedings{BredinPyannote2023,
  author={Hervé Bredin},
  title={{pyannote.audio 2.1 speaker diarization pipeline: principle, benchmark, and recipe}},
  year=2023,
  booktitle=interspeech,
}

@article{landiniBayesianHMMClustering2022,
  title = {Bayesian {{HMM}} Clustering of X-Vector Sequences ({{VBx}}) in Speaker Diarization: {{Theory}}, Implementation and Analysis on Standard Tasks},
  shorttitle = {Bayesian {{HMM}} Clustering of X-Vector Sequences ({{VBx}}) in Speaker Diarization},
  author = {Landini, Federico and Profant, J{\'a}n and Diez, Mireia and Burget, Luk{\'a}{\v s}},
  year = {2022},
  month = jan,
  journal = csl,
  volume = {71},
  pages = {101254},
  issn = {08852308},
  doi = {10.1016/j.csl.2021.101254}
}

@article{yoshiokaGeneralizationMultichannelLinear2012,
  title = {Generalization of Multi-Channel Linear Prediction Methods for Blind {{MIMO}} Impulse Response Shortening},
  author = {Yoshioka, Takuya and Nakatani, Tomohiro},
  year = {2012},
  journal = ieee-taslp,
  volume = {20},
  number = {10},
  pages = {2707--2720},
  doi = {10.1109/TASL.2012.2210879}
}

@inproceedings{fujitaEndEndNeuralSpeaker2019,
  author={Yusuke Fujita and Naoyuki Kanda and Shota Horiguchi and Kenji Nagamatsu and Shinji Watanabe},
  title={{End-to-End Neural Speaker Diarization with Permutation-free Objectives}},
  booktitle={Interspeech},
  pages={4300--4304},
  year=2019
}

@article{horiguchiEncoderDecoderBasedAttractors2022,
  title = {Encoder-{{Decoder Based Attractors}} for {{End-to-End Neural Diarization}}},
  author = {Horiguchi, Shota and Fujita, Yusuke and Watanabe, Shinji and Xue, Yawen and Garcia, Paola},
  year = {2022},
  journal = ieee-acm-taslp,
  volume = {30},
  primaryclass = {eess},
  pages = {1493--1507},
  issn = {2329-9290, 2329-9304},
  doi = {10.1109/TASLP.2022.3162080},
}

@inproceedings{snyderDeepNeuralNetwork2017,
  title = {Deep {{Neural Network Embeddings}} for {{Text-Independent Speaker Verification}}},
  booktitle = interspeech,
  author = {Snyder, David and {Garcia-Romero}, Daniel and Povey, Daniel and Khudanpur, Sanjeev},
  year = {2017},
  month = aug,
  pages = {999--1003},
  publisher = {ISCA},
  doi = {10.21437/Interspeech.2017-620}
}

@inproceedings{wangDiarizationLMSpeakerDiarization2024,
  title = {{{DiarizationLM}}: {{Speaker}} Diarization Post-Processing with Large Language Models},
  booktitle = interspeech,
  author = {Wang, Quan and Huang, Yiling and Zhao, Guanlong and Clark, Evan and Xia, Wei and Liao, Hank},
  year = {2024},
  pages = {3754--3758},
  issn = {2958-1796},
  doi = {10.21437/Interspeech.2024-209}
}

@misc{yinSpeakerLMEndtoEndVersatile2025,
  title={{SpeakerLM: End-to-End Versatile Speaker Diarization and Recognition with Multimodal Large Language Models}}, 
      author={Han Yin and Yafeng Chen and Chong Deng and Luyao Cheng and Hui Wang and Chao-Hong Tan and Qian Chen and Wen Wang and Xiangang Li},
      year={2025},
      eprint={2508.06372},
      archivePrefix={arXiv},
      primaryClass={cs.SD},
      url={https://arxiv.org/abs/2508.06372}, 
}

@inproceedings{mariotteASoBOAttentiveBeamformer2024,
  title = {{{ASoBO}}: {{Attentive}} Beamformer Selection for Distant Speaker Diarization in Meetings},
  booktitle = interspeech,
  author = {Mariotte, Th{\'e}o and Larcher, Anthony and Montr{\'e}sor, Silvio and Thomas, Jean-Hugh},
  year = {2024},
  pages = {1620--1624},
  issn = {2958-1796},
  doi = {10.21437/Interspeech.2024-917}
}

@inproceedings{pardoSpeakerDiarizationMultiple2006,
  title = {Speaker Diarization for Multiple Distant Microphone Meetings: Mixing Acoustic Features and Inter-Channel Time Differences},
  shorttitle = {Speaker Diarization for Multiple Distant Microphone Meetings},
  booktitle = interspeech,
  author = {Pardo, Jose M. and Anguera, Xavier and Wooters, Chuck},
  year = {2006},
  month = sep,
  pages = {paper 1337-Thu1A1O.5-0},
  doi = {10.21437/Interspeech.2006-570}
}

@inproceedings{cornellCHiME8DASRChallenge2024,
  title = {The {{CHiME-8 DASR}} Challenge for Generalizable and Array Agnostic Distant Automatic Speech Recognition and Diarization},
  booktitle = {8th International Workshop on Speech Processing in Everyday Environments},
  author = {Cornell, Samuele and Park, Tae Jin and Huang, He and Boeddeker, Christoph and Chang, Xuankai and Maciejewski, Matthew and Wiesner, Matthew S and Garcia, Paola and Watanabe, Shinji},
  year = {2024},
  pages = {1--6},
  doi = {10.21437/CHiME.2024-1}
}

@article{angueraAcousticBeamformingSpeaker2007,
  title = {Acoustic Beamforming for Speaker Diarization of Meetings},
  author = {Anguera, Xavier and Wooters, Chuck and Hernando, Javier},
  year = {2007},
  journal = ieee-taslp,
  volume = {15},
  number = {7},
  pages = {2011--2022},
  doi = {10.1109/TASL.2007.902460}
}

@inproceedings{scheiblerSurrogateSourceModel2021,
  title = {Surrogate {{Source Model Learning}} for {{Determined Source Separation}}},
  booktitle = icassp,
  author = {Scheibler, Robin and Togami, Masahito},
  year = {2021},
  month = jun,
  pages = {176--180},
  issn = {2379-190X},
  doi = {10.1109/ICASSP39728.2021.9414255}
}

@inproceedings{horiguchiMultiChannelEndtoEndNeural2022,
    title = {Multi-{{Channel End-to-End Neural Diarization}} with {{Distributed Microphones}}},
    author = {Horiguchi, Shota and Takashima, Yuki and Garcia, Paola and Watanabe, Shinji and Kawaguchi, Yohei},
    year = 2022,
    booktitle = ICASSP,
    pages={7332-7336}
}

@inproceedings{watanabe20b_chime,
  title = {{{CHiME-6}} Challenge: {{Tackling}} Multispeaker Speech Recognition for Unsegmented Recordings},
  booktitle = {6th International Workshop on Speech Processing in Everyday Environments ({{CHiME}} 2020)},
  author = {Watanabe, Shinji and Mandel, Michael and Barker, Jon and Vincent, Emmanuel and Arora, Ashish and Chang, Xuankai and Khudanpur, Sanjeev and Manohar, Vimal and Povey, Daniel and Raj, Desh and Snyder, David and Subramanian, Aswin Shanmugam and Trmal, Jan and Yair, Bar Ben and Boeddeker, Christoph and Ni, Zhaoheng and Fujita, Yusuke and Horiguchi, Shota and Kanda, Naoyuki and Yoshioka, Takuya and Ryant, Neville},
  year = 2020,
  pages = {1--7},
  doi = {10.21437/CHiME.2020-1}
}

@inproceedings{yu2022alimeeting,
  title = {{{M2MeT}}: {{The ICASSP}} 2022 Multi-Channel Multi-Party Meeting Transcription Challenge},
  booktitle = {Proceedings of the {{IEEE}} International Conference on Acoustics, Speech and Signal Processing},
  author = {Yu, Fan and Zhang, Shiliang and Fu, Yihui and Xie, Lei and Zheng, Siqi and Du, Zhihao and Huang, Weilong and Guo, Pengcheng and Yan, Zhijie and Ma, Bin and Xu, Xin and Bu, Hui},
  year = 2022,
  eprint = {2110.07393},
  primaryclass = {eess.AS},
  pages = {6167--6171},
  publisher = {IEEE},
  archiveprefix = {arXiv}
}

@inproceedings{horiguchi_asru2025,
    author = {Horiguchi, Shota and Tawara, Naohiro and Ashihara, Takanori and Ando, Atsushi and Delcroix, Marc},
    title = {Can We Really Repurpose Multi-Speaker ASR Corpus for Speaker Diarization?},
    booktitle={IEEE Automatic Speech Recognition and Understanding Workshop (ASRU)},
    year = {2025},
    month = {Dec},
}

@inproceedings{sumuraJointAudioSource2024,
  title = {Joint Audio Source Localization and Separation with Distributed Microphone Arrays Based on Spatially-Regularized Multichannel {{NMF}}},
  booktitle = {2024 18th International Workshop on Acoustic Signal Enhancement ({{IWAENC}})},
  author = {Sumura, Yoshiaki and Carlo, Diego Di and Arie Nugraha, Aditya and Bando, Yoshiaki and Yoshii, Kazuyoshi},
  year = 2024,
  pages = {145--149},
  doi = {10.1109/IWAENC61483.2024.10694042}
}

@misc{mao2026neuralmultichanneldistantspeaker,
      title={Neural Multichannel Distant Speaker Diarization and Source Separation with Beta Speaker Activity Prior}, 
      author={Sicheng Mao and Mathieu Fontaine and Anthony Larcher and Roland Badeau},
      year={2026},
      eprint={2608.28661},
      archivePrefix={arXiv},
      primaryClass={cs.SD},
      url={https://arxiv.org/abs/2608.28661}, 
}

@inproceedings{kingmaSemisupervisedLearningDeep2014,
  title = {Semi-Supervised Learning with Deep Generative Models},
  author = {Kingma, Diederik P and Rezende, Danilo J and Mohamed, Shakir and Welling, Max},
  year = 2014,
  volume = {27},
  booktitle = {{{NIPS}}}
}
\end{document}